\documentclass{PoS}
\usepackage{float}
\usepackage{wrapfig}

\title{Calibration and operation of SiPM-based cameras for gamma-ray astronomy in presence of high night-sky light}

\ShortTitle{Performances of the SST-1M camera}

\newcommand{\krakow}{Krak\'ow{}}
\author{\speaker{C.~Alispach}$^a$, I.~Al~Samarai$^a$, M.~Balbo$^b$, A.~Barbano$^a$, V.~Beshley$^o$, A.~Biland$^c$, J.~Blazek$^i$, J.~B{\l}ocki$^d$, J.~Borkowski$^h$, T.~Bulik$^e$, F.~Cadoux$^a$, L.~Chytka$^l$, V.~Coco$^a$, N.~De Angelis$^a$, D.~della Volpe$^a$, Y.~Favre$^a$, T.~Gieras$^d$, M.~Grudzi{\'n}ska$^e$, P.~Hamal$^l$, M.~Heller$^a$, M.~Hrabovsky$^l$, J.~Jury\v{s}ek$^i,l$, J.~Kasperek$^k$, K.~Koncewicz$^d$, A.~Kotarba$^d$, E.~Lyard$^b$, E.~Mach$^d$, D.~Mandat$^i$, S.~Michal$^l$, J.~Micha{\l}owski$^d$, R.~Moderski$^h$, T.~Montaruli$^a$, A.~Nagai$^a$, D.~Neise$^c$, J.~Niemiec$^d$, T.R.S.~Njoh~Ekoume$^a$, M.~Ostrowski$^f$, M.~Palatka$^i$, P.~Pa{\'s}ko$^g$, H.~Przybilski$^d$, M.~Pech$^i$, B.~Pilszyk$^d$, P.~Rajda$^k$,  Y.~Renier$^a$, P.~Rozwadowski$^e$, P.~Schovanek$^i$, K.~Seweryn$^g$,  V.~Sliusar$^b$, D.~Smakulska$^d$, D.~Sobczy\'{n}ska$^n$, {\L}.~Stawarz$^f$,J.~\'{S}wierblewski$^d$, P.~\'{S}wierk$^d$,P.~Travnicek$^i$, I.~Troyano Pujadas$^a$, R.~Walter$^b$, M.~Wiecek$^d$, A.~Zagda\'{n}ski$^f$, K.~Zi{\c e}tara$^f$, for the SST-1M project\\
\llap{$^a$}\textit{DPNC - Universit\'e de Gen\`eve, 24 Quai Ernest Ansermet, CH-1211 Gen\`eve,  Switzerland}\\
\llap{$^b$}\textit{D\'epartement d'Astronomie, Universit\'e de Gen\`eve, Chemin d'Ecogia 16, CH-1290 Versoix, Switzerland}\\
\llap{$^c$}\textit{ETH Zurich, Institute for Particle Physics and Astrophysics, Otto-Stern-Weg 5, 8093 Zurich, Switzerland}\\
\llap{$^d$}\textit{Institute of Nuclear Physics Polish Academy of Sciences, PL-31342 Krakow, Poland}\\
\llap{$^e$}\textit{Astronomical Observatory, University of Warsaw, Al. Ujazdowskie 4, 00-478 Warsaw, Poland}\\
\llap{$^f$}\textit{Astronomical Observatory, Jagiellonian University, ul. Orla 171, 30-244 \krakow, Poland}\\
\llap{$^g$}\textit{Centrum Bada{\'n} Kosmicznych Polskiej Akademii Nauk,  18a Bartycka str., 00-716 Warsaw, Poland }\\
\llap{$^h$}\textit{Nicolaus Copernicus Astronomical Center, Polish Academy of Sciences,  ul. Bartycka 18, 00-716 Warsaw, Poland }\\
\llap{$^k$} \textit{AGH University of Science and Technology, al.Mickiewicza 30,  30-059 \krakow, Poland}\\
\llap{$^i$} \textit{FZU - Institute of Physics of the Czech Academy of Sciences, 17. listopadu 50, Olomouc \& Na Slovance 2, Prague, Czech Republic.}\\
\llap{$^l$} \textit{Palacky University Olomouc, Faculty of Science, RCPTM, 17. listopadu 50, Olomouc, Czech Republic.}\\
\llap{$^n$} \textit{Department of Astrophysics, University of {\L}\'od\'z, ul. Pomorska 149/153, 90-236 {\L}\'od\'z, Poland }\\
\llap{$^o$} \textit{Pidstryhach Institute for Applied Problems of Mechanics and Mathematics, National Academy of Sciences of Ukraine, 3-b Naukova St., 79060, Lviv, Ukraine} \\
E-mail: \email{cyril.alispach@unige.ch}}

\abstract{
The next generation of Cherenkov telescope cameras feature SiPM, which can guarantee excellent performance and allow for observation also under moonlight. We present the calibration and performance  of a 1296-pixel SiPM camera prototype for gamma-ray astronomy.
}

\FullConference{36th International Cosmic Ray Conference -ICRC2019-\\
		July 24th - August 1st, 2019\\
		Madison, WI, U.S.A.}

\begin{document}

\section{Introduction}

Ground-based gamma-ray telescopes rely on the imaging atmospheric Cherenkov technique (IACT), which uses the Cherenkov photons emitted by the charged particles of an extensive air shower (EAS) in order to reconstruct the arrival directions and energies of the initial photon or charged particle. The Cherenkov photons are detected on the ground with telescopes equipped with large mirrors. The image obtained by the telescope camera is then used to determine the type of the primary particle inducing the EAS, its energy and incoming direction. \\

Nowadays Cherenkov telescope cameras~\cite{HESS-camera}~\cite{MAGIC-camera} use most often Photo Multiplier Tubes (PMTs). The PMTs offer great sensitivity in the UV band and enable the detection of the faint and short lived flashes of Cherenkov light produced in the atmosphere. However, the PMTs have a shortened life time when exposed to background light. For this reason most current gamma-ray observations are performed in the darkest conditions while avoiding nights of high-Moon intensity. This reduces the duty cycle of the telescopes. \\

To increase the duty cycle of Cherenkov telescopes, Silicon Photo Multiplier (SiPM) based Cherenkov camera have been built and operated by the FACT collaboration~\cite{FACT}. The FACT collaboration showed that SiPM-based cameras can be operated without damage in high night-sky background conditions. The recent progress in SiPM technology, such as the increased photodetection efficiency in the UV band, the improved photoelectron resolution and the reduction in optical crosstalk, have allowed SiPMs to be considered  as a replacement for PMTs in next-generation Cherenkov cameras~\cite{Otte:2016aaw}. \\

The performances of the instrument presented here are compared to the highly demanding requirements set by the next-generation ground-based gamma-ray observatory: the Cherenkov Telescope Array (CTA)~\cite{CTA-concept}.
 
\section{The SST-1M camera and its Calibration Test Setup (CTS)}

The SST-1M camera~\cite{CameraPaperHeller2017} is a SiPM-based Cherenkov camera for the SST-1M telescope~\cite{SST1M-project-ICRC19}. It consists of 1296 hexagonal SiPMs from Hammamatsu (S10943-3739(X)) connected to a front-end electronics for signal shaping and amplification designed at the University of Geneva~\cite{Frontend2016}. The front-end signals are recorded by the trigger and readout system DigiCam~\cite{Rajdak2015}. Continuous digitization with 12-bit Flash Analog to Digital Converters (FADCs) of the signal is performed at a sampling frequency of 250~MHz. A ring buffer keeps the data within the system while continuing the observations, allowing dead time free operations. \\

Calibration of the SiPMs is performed with a dedicated  Camera Test Setup (CTS). A description of this calibration setup can be found in~\cite{Alispach-ICRC17} which has since then been upgraded to an array of 1296 pairs of LED that covers the entire camera field of view. Each pair is placed in front of a camera pixel and consists of a pulsed and a continuously operated LED. The pulsed LED (AC LED) emulates the prompt Cherenkov signal while the continuous LED (DC LED) emulates the night-sky background (NSB). The CTS is now fully controllable by the telescope control software~\cite{Sliusar2017-ACS} and does not require an external pulse generator. The pulse generation is carried out by one of the DigiCam microcrates which makes the CTS fully autonomous and usable on any observation site, only requiring a 230~V power plug.\\

This calibration system was developed in view of a mass production of the cameras (70 pieces at a rate of two cameras per month). The whole procedure described in the following sections would require a couple of days for the data acquisition and another day for the data analysis. The data analysis, presented in the following, was performed with a dedicated Python software: \textit{digicampipe}~\cite{digicampipe} based on the core functionaries of the proposed reconstruction pipeline for CTA: \textit{ctapipe}~\cite{ctapipe-ICRC19} \\

\section{Camera performances in the presence of night-sky background}\label{sec:key_perf}

In order to fully characterize each pixel, the camera pixels are illuminated with the CTS mounted in front of it. The data sample used for calibration consists of a scan of pulsed light from 0 to $\sim 10$~k photons and continuous light from 0 to $\sim 1$~GHz photoelectron rate per pixel. For each light level a set of 10~k waveforms at 500~Hz are acquired. The subset without background continuous light is used to calibrate the pulsed LEDs using the photo counting capabilities of SiPM as performed in~\cite{CameraPaperHeller2017}. It is also used to extract key SiPM parameters such as gain, dark count rate, gain smearing and optical crosstalk for each pixel. The subset without pulsed light is used to calibrate the continuous LEDs from the pedestal shift. \\

Once the LEDs are calibrated and the SiPM characterized the rest of the dataset is used to measure the time and charge resolution of the camera. The reconstructed time and charge of an EAS play a crucial role in the reconstruction of the impact parameter and the primary particle energy. Therefore time and charge resolution are both instrument response functions directly linked to the telescope performance on gamma/hadron separation, angular resolution and energy resolution. The Monte Carlo evaluation of the telescope performances can be found in~\cite{SST1M-monte-carlo-ICRC19}. In the following we will present the time and charge resolution of the SST-1M camera. 

\subsection{Time resolution}

% The full analysis described below has been made available in digicampipe~\cite{digicampipe} and the script \verb|digicampipe/bash/time_resolution.sh| can reproduce all the plots shown in this section.

The flashes of the CTS were triggered by the digital readout itself which ensures the synchronicity between the readout and the light pulse. Doing so, the light arrives every time at the same position in the readout window. To measure the time offset $\delta t$, we use the charge ratio $a$ between the pulse template function $f$ and the $N$ waveform samples $x_i$. This ratio $a$ is calculated using the sum over 8 consecutive samples (3 before the maximum and 4 after) individually for the samples and for the template:
\begin{equation}
    a(\delta t) = \frac{\sum\limits_{i=-3}^{4}  x_{i - i_{max}}} {\sum\limits_{j=-3}^{4}  f(t_j+\delta t-t_{max})}
\end{equation}

The $\chi^2$ giving the agreement between the pulse and the scaled template with an offset $\delta t$ is calculated as:
\begin{equation}
    \chi^2(\delta t) = \frac{1}{N} \sum_{i=1}^{N} \frac{(x_i - a(\delta t) f(t_i+\delta t))^2}{\sigma_{f(t_i+\delta t)}^2 + \sigma_e^2}
\end{equation}
where:
\begin{itemize}
    % \item $N$ is the number of samples
    % \item $x_i$ is the amplitude of the pulse at the sample $i$
    \item $f(t_i+\delta t)$ and $\sigma_{f(t_i+\delta t)}$ are respectively the amplitude and the uncertainty of the template for the sample $i$ and an offset $\delta t$
    \item $\sigma_e$ is the electronic noise per time slice
\end{itemize}

The $\chi^2$ is calculated for a range of offsets with 0.1~ns steps (as smaller steps did not improve the results), for each pixel and each flash. The time offset $\delta s$ corresponding to the lowest $\chi^2$ is chosen as the measurement of the reconstructed time offset for that flash. The time offset (respectively the time resolution) for each pixel is obtained as the mean value (respectively the standard deviation) over the sample. Only flashes where the measured amplitude is above 3.5~p.e. are used.

% This measurement is repeated for several pulse LED amplitudes and the corresponding number of p.e. is derived using LED calibrations curves (see fig.~\ref{fig:leds}). These pulsed amplitude scans were done both without any continuous light and with a NSB level of about to 125MHz (DC level of 290).

\begin{figure}[H]
\begin{center}
\includegraphics[width=0.49\textwidth]{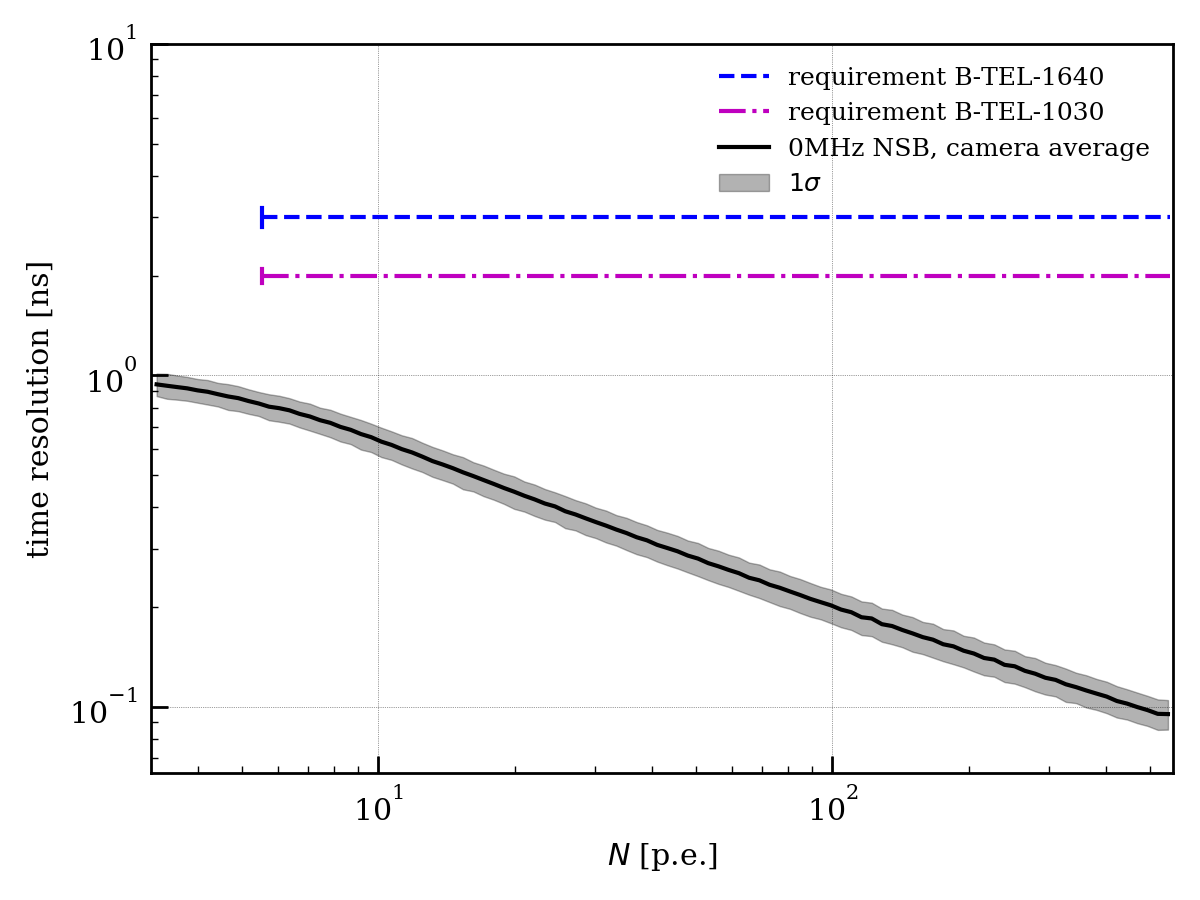}
\includegraphics[width=0.49\textwidth]{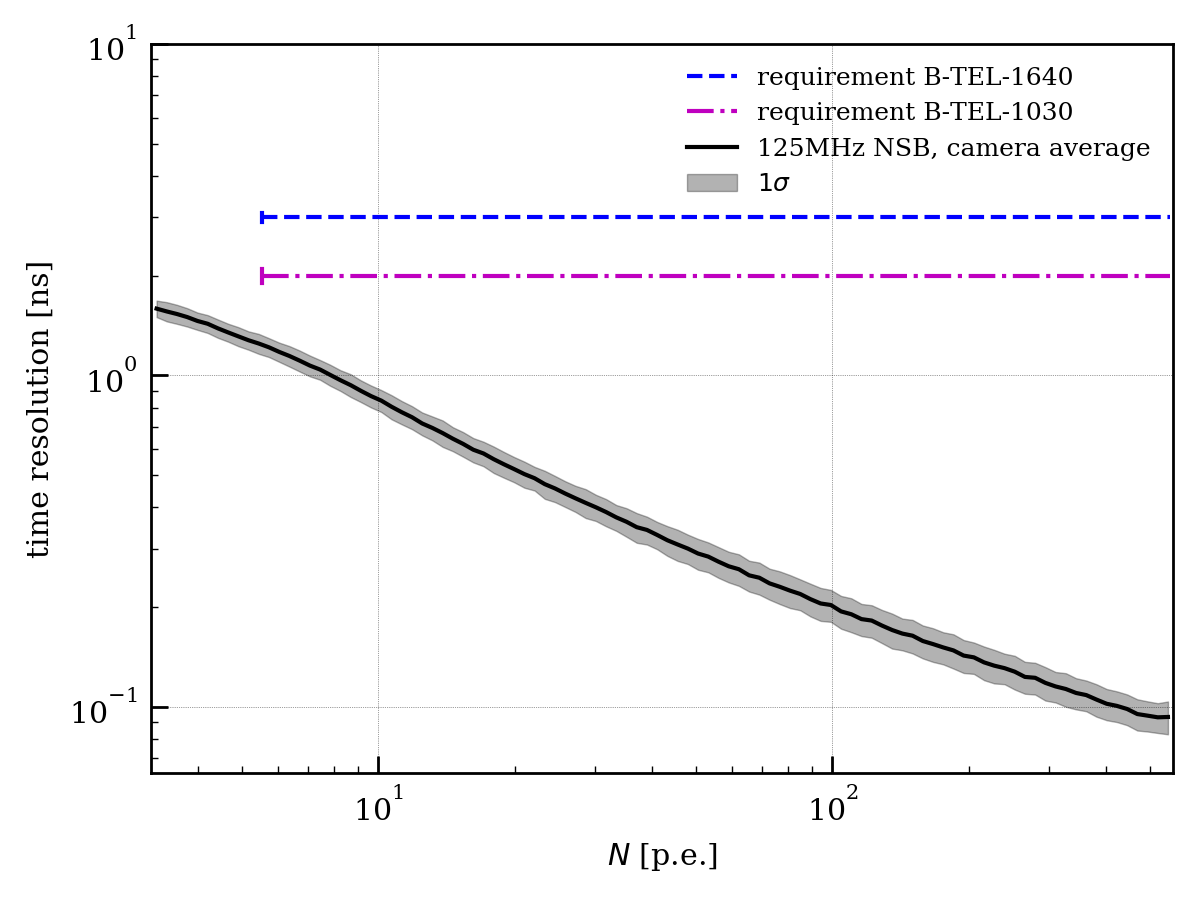}
\end{center}
\caption{Time resolution as a function of the reconstructed charge in dark conditions (left) and with 125~MHz NSB equivalent (right). The time resolution is compared to the current CTA requirements for a 125~MHz NSB level.}\label{fig:time_resol}
\end{figure}

Fig. \ref{fig:time_resol} shows for the whole camera the evolution of the timing resolution with the charge for two NSB levels. We see that without any NSB, the resolution is below 1~ns and reaches 0.1~ns at 400 p.e. With 125 MHz NSB, the resolution is mainly affected below 50 p.e. and goes above 1~ns only for pulse amplitude below 7 p.e. (~30 photons).

%Fig. \ref{fig:time_offset} shows the evolution of the timing offset with the charge for these two NSB levels for the whole camera. This dependency  comes from the LED pulse and is similar for all LEDs. 

%\begin{figure}[H]
%\begin{center}
%\includegraphics[width=0.49\textwidth]{images/time_offset_dark.png} \hfill
%\includegraphics[width=0.49\textwidth]{images/time_offset_dc290.png}
%\end{center}
%\caption{Time offset function of charge in dark conditions (left) and with 125~MHz NSB equivalent (right). For each pixel, the offset interpolated at 100 p.e. is taken as reference.}\label{fig:time_offset}
%\end{figure}

% Figure \ref{fig:time_camera} shows the the timing offset and timing resolution for each pixels on the camera. As the relative FACs of DigiCam have been measured to be below 1~fs, the measured offset non-uniformity must be coming from the CTS. One can also notice a strong correlation between the offsets and the resolution, which indicates the time resolution is mostly impacted by the jitter on the light emission from the LEDs.

%\begin{figure}[H]
%\begin{center}
%\includegraphics[width=0.49\textwidth]{images/time_offset_camera.png}
%\includegraphics[width=0.49\textwidth]{images/time_resolution_camera.png}
%\end{center}
%\caption{Time offset and time resolution without NSB and 100 p.e %flash.}\label{fig:time_camera}
%\end{figure}

\subsection{Charge resolution}

% The charge resolution measures the resolutions of the camera pixels and the data acquisition chain to a Poisson light source. The charge resolution is plays a crucial role in the energy resolution of the telescope. In the field of gamma-ray astronomy the resolution in energy is quiet poor.\\

The data analysis performed here is similar to the one described in~\cite{CameraPaperHeller2017}. Compared to the previous results obtained in~\cite{CameraPaperHeller2017}, the measurements were performed for each of the 1296 camera pixels and are expressed in photon units rather than photoelectrons. They are compared to the current CTA requirements.

The charge resolution is defined as the ratio between the variance and mean of the reconstructed charge in units of photons.
\begin{equation}
    CR = \frac{Var(N_{\gamma})}{\mathbb{E}(N_{\gamma})}
    \label{eq:charge_reso}
\end{equation}
The variance and mean are computed with the sample standard deviation and mean. The charge resolution presented here takes into account: Poisson fluctuations of the LED, electronic noise from the photodetection plane, SiPM gain smearing, optical crosstalk, precision of the computed baseline and non linearity of the amplification chain. \\

\begin{figure}[H]
\begin{center}
\includegraphics[width=0.49\textwidth]{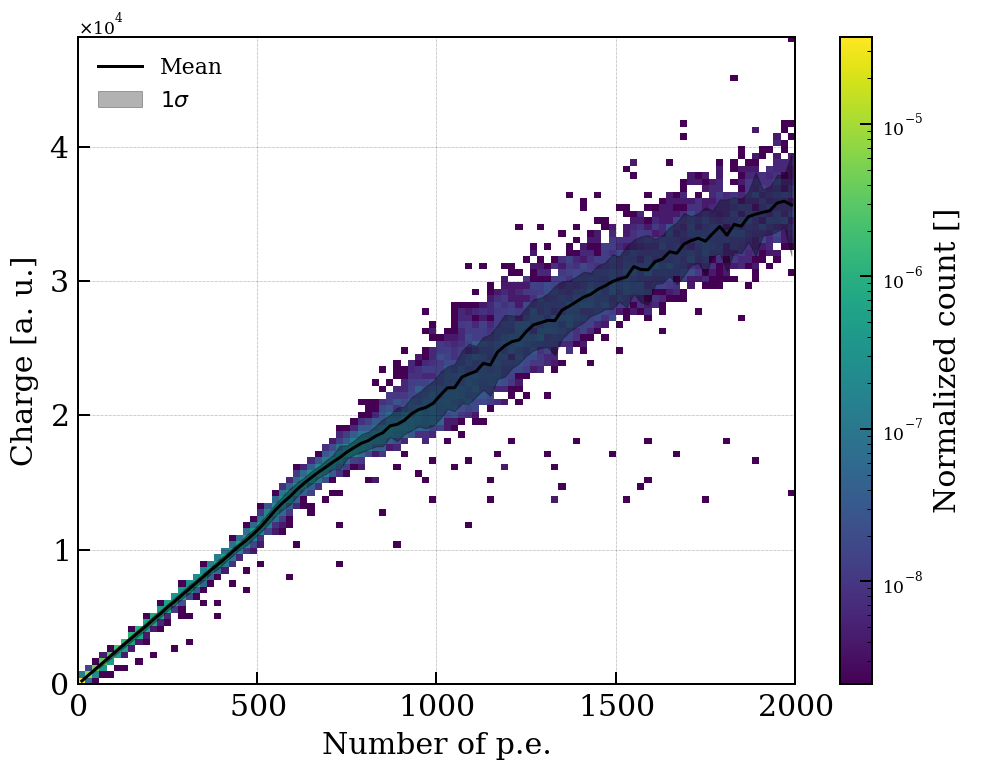}
\includegraphics[width=0.49\textwidth]{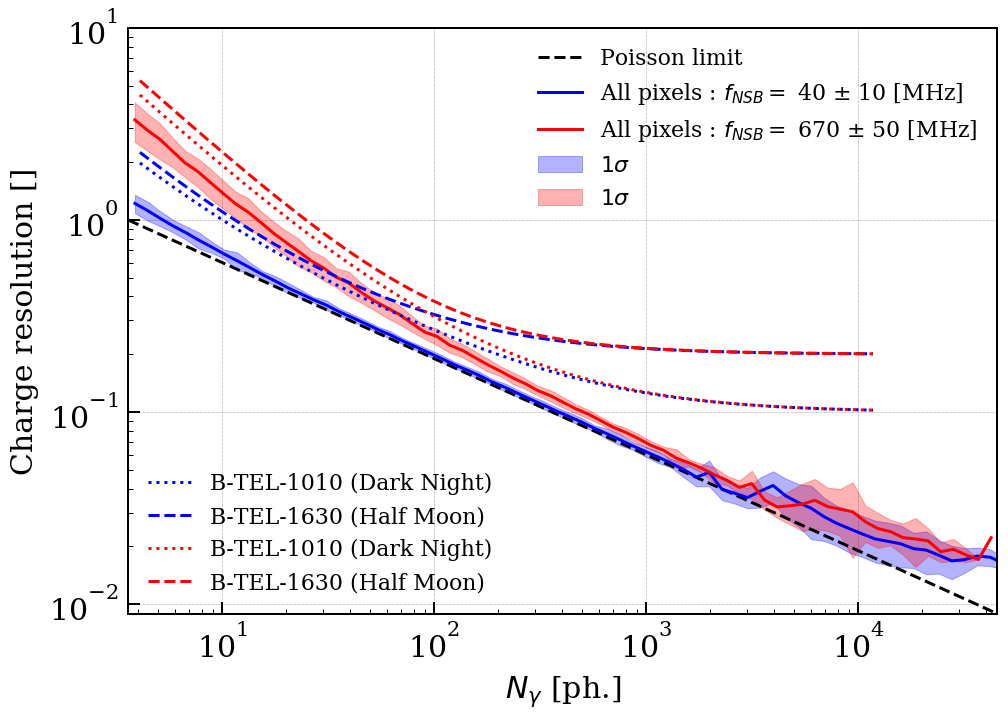}
\end{center}
\caption{Left: Reconstructed charge as a function of the true number of photoelectrons for all camera pixels. Right: Charge resolution as a function of the true number of photons for different night-sky background levels. The solid lines represents the average charge resolution of the camera and the contoured band its 1-sigma deviation among camera pixels.}\label{fig:charge_resolution}
\end{figure}

The readout chain starts to saturate at high illumination due to the pre-amplifier saturation. To cope with saturation effects, a look-up table of reconstructed charge (waveform integral) as a function of the true number of photoelectrons is used (see Fig. \ref{fig:charge_resolution}) for all camera pixels.  The number of photons is then assessed by correcting for the optical efficiency of the photodectection plane. The correction of the number of photoelectrons induced by a voltage drop from night-sky background are considered as described in~\cite{VdropHeller}. \\

The charge resolution is presented in Fig.~\ref{fig:charge_resolution}. The solid lines denote the average resolution over the camera pixels and the contoured area represents its $1\sigma$ deviation. The theoretical Poisson limit (fluctuation of the light source only) is given in black. The charge resolution is given for two distinct night-sky background rates per pixel (in photoelectrons) corresponding to clear sky conditions (40~MHz) and to a half-Moon night (670~MHz) in Paranal, Chile. The corresponding CTA requirements for the small-sized telescopes are also drawn which correspond to two data processing levels (dotted and dashed lines). As shown the SST-1M camera is compliant with the CTA requirements. \\

\section{On-site monitoring}\label{sec:on-site}

The SST-1M camera has been subject to extensive integration and commissioning to the telescope prototype structure in IFJ Krakow during autumn 2018. A few nights were dedicated to observations of the Crab Nebula and its first analysis is described in~\cite{SST1M-monte-carlo-ICRC19}. Prior to and during observations, calibration runs are performed to ensure the stability and operability of the camera. The safety and operability of the various camera components (such as temperatures of the SiPMs, readout rates, trigger rates, etc.) are monitored via a slow control link at 2~Hz. In this section we present the dark runs and trigger monitoring performed on site.

\subsection{Dark run}\label{sec:dark_runs}

\begin{figure}[H]
    \centering
    \includegraphics[width=0.49\textwidth]{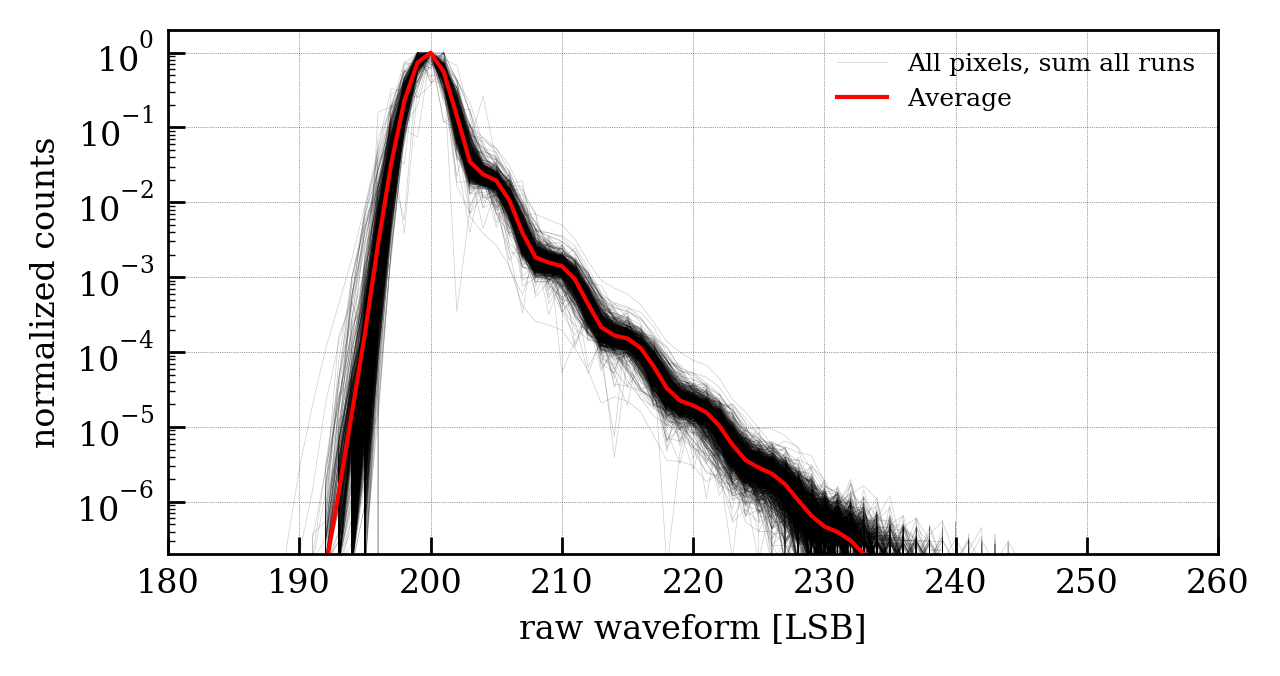} \hfill
    \includegraphics[width=0.49\textwidth]{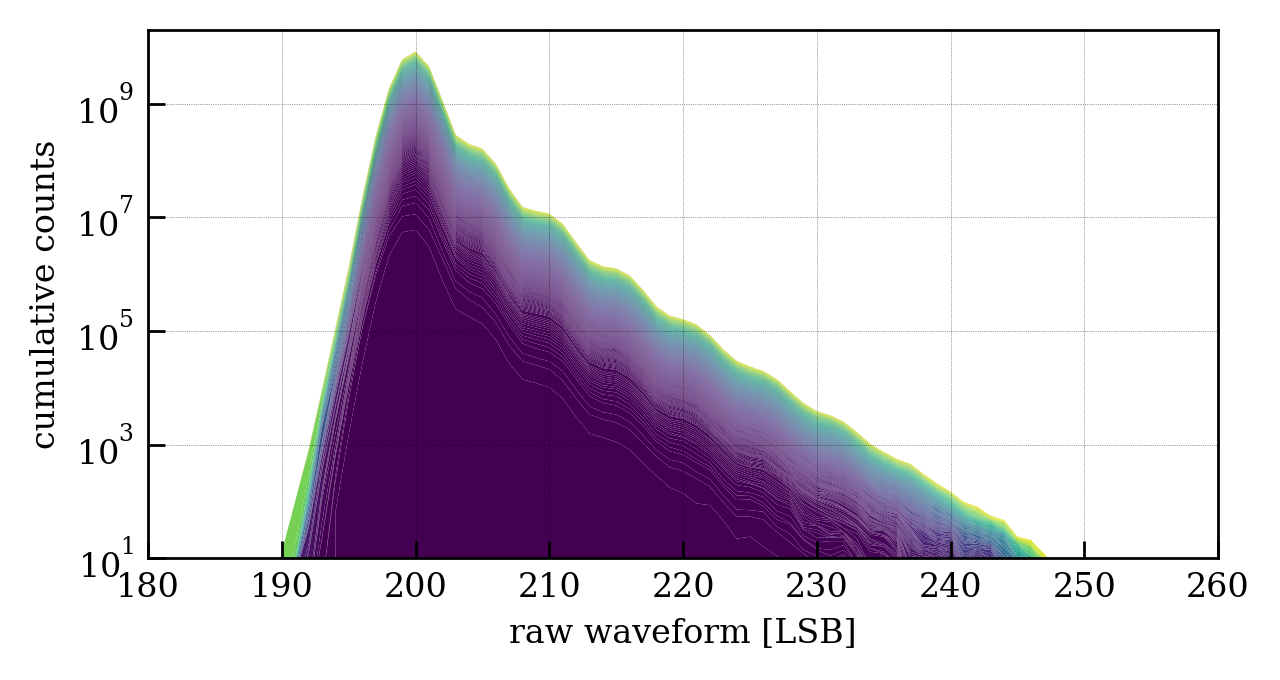}
    \caption{Left: Normalized distribution of the projection of the raw waveform in LSB for all the pixels and for one run. Right: Merged distribution of all pixels and all nights.}
    \label{fig:dark_spe}
\end{figure}

Before each night of observation, while the lid is closed, a dark run is acquired. Ten thousand waveforms of 50 samples for each pixel are registered. Their first use is to measure the baseline without any night-sky background to determine later the baseline shift per pixel (later used to compute the night-sky background level). Their second use is to monitor the sensor parameters evolution, e.g. gain, optical crosstalk and dark count rate. Fig.~\ref{fig:dark_spe} shows the projection of the raw waveform in Least Significant Bit (LSB) for each camera pixel. Given that, the first five photoelectron peaks are easily identifiable within a dark run, the distribution shown on  Fig.~\ref{fig:dark_spe} is enough to extract with good precision the sensor parameters on a per night basis. Additionally, merging all the dark runs and/or merging all the pixels allows the necessary statistics to be accumulated to correct the optical crosstalk modelization as shown in~\cite{FACT}.

\subsection{Trigger rate}

The trigger logic is based on the clustering of neighboring signals in the camera pixels. The internal trigger of DigiCam sums the signal of each of the three neighboring pixels of the camera into so-called trigger patches. Each of the 432 patches are then summed with its neighboring six patches to form a trigger cluster. An event is triggered when at least one of the 432-cluster waveform signals passes the threshold. This trigger topology allows the rate of false triggers induced by night-sky background fluctuations to be reduced while capturing EAS events which illuminate simultaneously groups of nearby pixels. \\

\begin{figure}[H]
    \centering
    \includegraphics[width=0.49\textwidth]{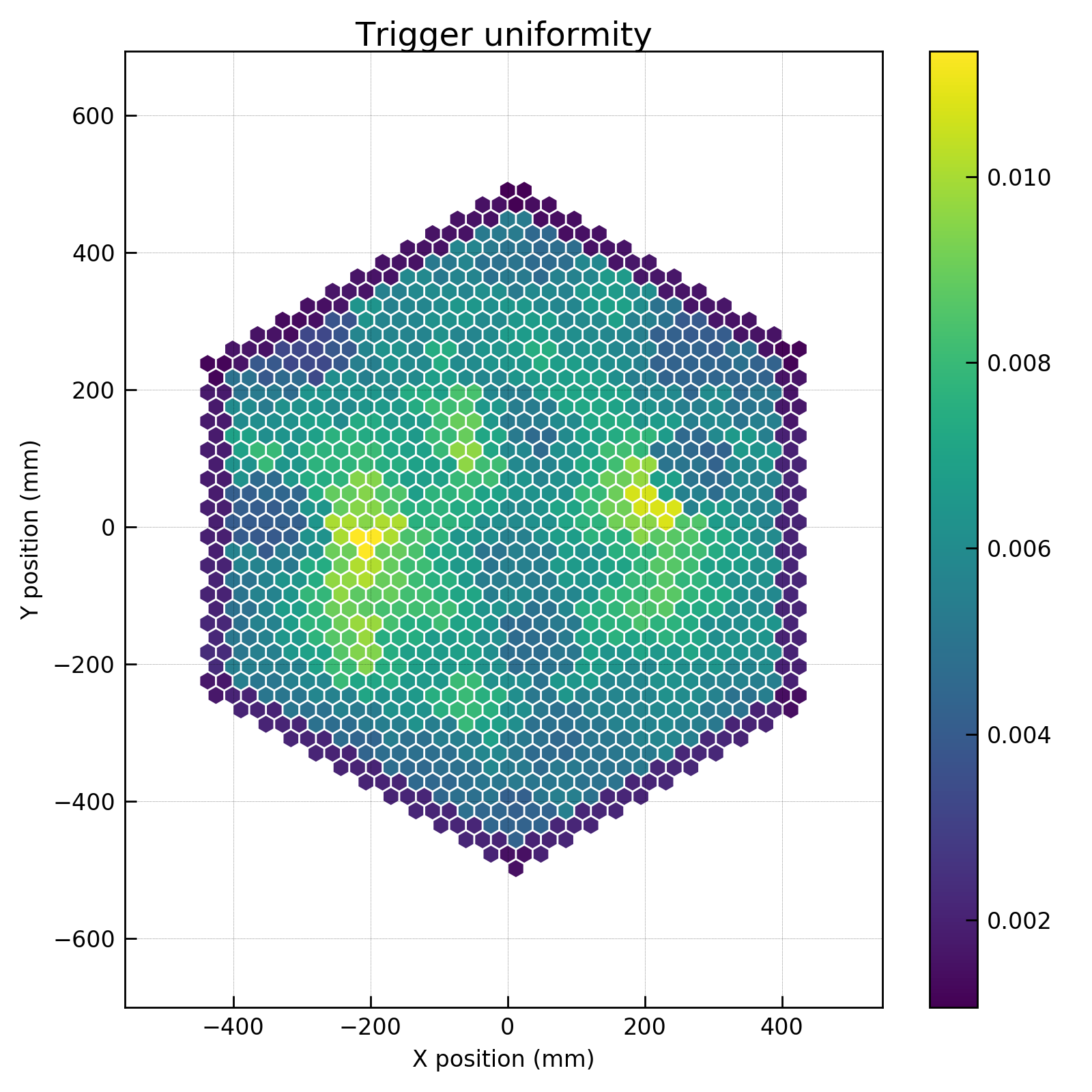}
    \caption{Trigger uniformity of the EAS shower events during the observation night of the 12$^{th}$ october 2018}
    \label{fig:trigger_uniformity}
\end{figure}

During the 2018 operation of the telescope, high night-sky background conditions of 600~MHz on average per pixel were observed~\cite{SST1M-project-ICRC19}. On top of the high night-sky background level, external parasitic light from the surrounding inhabited area perturbed the observations. To limit the influence of such background during science runs, the fully digital trigger logic allows trigger clusters to be disabled such that they would not trigger the readout of the full camera. This is needed when a bright continuous light source falls into the field of view of a pixel or a group of pixels. Alternatively, the signal can be clipped at the pixel. Even though both SiPM and FADC gains have been equalized providing a 3\% rms spread over the camera gain, the variation in optical efficiencies were not corrected. For instance, the transmissivity of the entrance window and of the light guides are so far not compensated. Once determined using an external light source (e.g. Flasher), the correction factors can be compensated as the trigger threshold can be set at the patch level. This has the advantage to maintain all SiPMs at the same working point and avoids the need to compensate with the bias voltage which would lead to sensors working with different characteristics (noise, gain, optical crosstalk, dark count rate, etc.). Fig.~\ref{fig:trigger_uniformity} shows the fractional contribution of each trigger cluster to the camera internal trigger for one night of observation during the 2018 observation campaign. The non-uniformities observed here are mostly related to different NSB levels across the camera FoV.

\section{Conclusion}

The SST-1M camera shows reliable performance even in the presence of high night-sky background conditions. Its performance matches the high requirements set by the next generation of gamma-ray instruments. 

\acknowledgments

This work was supported by the grant Nr. DIR/WK/2017/12 from the Polish Ministry of Science and Higher Education. We greatly acknowledge financial support form the Swiss State Secretariat for Education Research and Innovation SERI. The work is supported by the projects of Ministry of Education, Youth and Sports: MEYS LM2015046, LTT17006 and EU/MEYS CZ.02.1.01/0.0/0.0/16\_013/0001403, Czech Republic.
\bibliographystyle{JHEP} % We choose the "plain" reference style
\bibliography{paper_pos} % Entries are in the "refs.bib" file

\end{document}